# Does thiosemicarbazide lead nitrate (TSLN) crystal exist?

Royle Fernandes[a], Bikshandarkoil R. Srinivasan[b]
[a]Department of Chemistry, Smt Parvatibai Chowgule College of Arts & Science, Margao, Goa 403 602 Email: roylefernandes43@gmail.com
[b]Department of Chemistry, Goa University, Goa 403206, INDIA
Email: srini@unigoa.ac.in

**Abstract**

The authors of a recent paper (Optik 125 (2014) 2022-2025) claim to have grown a so called thiosemicarbazide lead nitrate (TSLN) crystal by the slow evaporation method. In this comment we prove that TSLN is actually thiosemicarbazide.

**Comment**

During the course of a literature survey we came across the title paper by Shakila and Kalainathan [1] reporting in the abstract "*Single crystal XRD study revealed that materials crystallized with triclinic crystal structure and its belongs to centrosymmetric space group Cc and powder X-ray diffraction(XRD) is to confirm the crystalline nature of the crystal*...", In view of a strange space group assignment in the triclinic crystal system this paper attracted our attention and was taken up for scrutiny for verification of the claim. A perusal of the paper reveals that the authors have grown crystals of a so called thiosemicarbazide lead nitrate, by the slow evaporation solution method. It is noted that in addition to an unusual name viz. thiosemicarbazide lead nitrate, not in accordance with IUPAC nomenclature, the crystals are abbreviated by a strange code TSLN.

Under the heading "*Single crystal X-ray diffraction and fundamental parameters*' the authors report the space group Cc but the cell parameters given are for a triclinic cell and further claim that their cell is in agreement with the reported cell in Ref. 7 of the commented paper [2]. The Cc space group cannot be a typographic error because in the triclinic system the only space groups are *P1* and *P-1*. A quick look of Reference 7 in the title paper (which is a M.Sc. Dissertaion) helps to resolve the problem of the strange triclinic space group. The unit cell parameters reported by the authors of the title paper appear to have been copied from page 63 of the dissertation (ref.2 in the present comment) which reports the unit cell measured at 100 K, because all the values including the esd's (a, b, c, alpha, beta, gamma) are exactly the same. The copying is so perfect that the authors copied the space group Cc which is also wrongly mentioned. Since no such cell (with the mentioned parameters is reported for a Cd-thiosemicarbazide compound in the Cambridge Data Base, we opine that the authors probably copied data from an unreliable source. It is noted that the authors do not mention any temperature for cell measurement. We find it truly remarkable that two unit cells measured in two different institutions are identical. As it is meaningless to refine in Cc space group a triclinic cell, we believe no single crystal X-ray measurement was actually done as is being claimed.

It is not clear why the authors decided to grow crystals of this so called TSLN which in their opinion was already a known crystal. Unfortunately, the other characterization data for TSLN for example EDAX, IR and UV-Vis spectra do not offer any valid scientific proof for the claimed composition. On the contrary the IR spectrum actually appears to be that of pure thiosemicarbazide. Recently we have shown that some metal-thiosemicarbazide compounds are actually thiosemicarbazide [3]. In order to verify the claim of preparation of a thiosemicarbazide containing lead nitrate in the lead nitrate-thiosemicarbaide system, we reinvestigated the crystal growth of TSLN under the same conditions as reported by Shakila and Kalainathan [1]. The product crystal which separated from the crystal growth medium turned out to be pure thiosemicarbazide as confirmed by its IR spectrum and melting point. Our experiment thus indicates that at RT, there is no chemical reaction between lead nitrate and thiosemicarbazde in water and the less soluble thiosemicarbazide fractionally crystallizes while the more soluble lead nitrate remains in solution. Hence it is not surprising that the IR spectrum reported in the commented paper matches with that of thiosemicarbazide which the authors failed to recognise. It is to be noted that the same research group had earlier reported to have made crystals of a so called urea thiosemicarbazone monohydrate by reaction of urea with thiosemmicarbazide. This has already been proved to be the less soluble thiosemicarbazide [1]. Although it is not clear why thiosemicarbazide is reported under different names, it can be concluded that TSLN is a dubious crystal.

## References

1. K. Shakila, S. Kalainathan, Growth and characterization of semiorganic crystal:Thiosemicarbazide lead nitrate (TSLN), Optik 125 (2014) 2022-2026, http://dx.doi.org/10.1016/j.ijleo.2013.08.047

2. http://uzspace.uzulu.ac.za/bitstream/handle/10530/273/Cadmium%20and%20lead%20thiosemicarbazide%20complexes%20-%20Mlondo%20SN.pdf?sequence=1

[3] B.R. Srinivasan, N. Keerthika, Reinvestigation of crystal growth of thiosemicarbazide potassium chloride and thiosemicarbazide lithium chloride, *Optik* 125 4807-4809 (2014). http://dx.doi.org/10.1016/j.ijleo.2014.04.054

[4] B.R. Srinivasan, P. Raghavaiah, V.S. Nadkarni, Reinvestigation of growth of urea thiosemicar-bazone monohydrate crystal, *Spectrochimica Acta* A112 84-89 (2013). http://dx.doi.org/10.1016/j.saa.2013.04.026